\pdfoutput=1

\documentclass[a4paper,11pt]{article}
\usepackage{jcappub} 
\usepackage{color}
\usepackage{xcolor}
\usepackage{amsmath}
\usepackage{amsfonts}
\usepackage{graphicx} 
\usepackage{amssymb}
\usepackage{epstopdf}
\usepackage{url}
\usepackage{bm}
\usepackage{hhline}
\usepackage{subfigure}
\usepackage{multirow}
\usepackage{epsfig}
\usepackage{varioref}
\usepackage{amssymb}
\usepackage{comment} 
\usepackage{multirow}
\usepackage{mathrsfs}
\usepackage{mathtools}
\usepackage{amsthm}
\usepackage{hyperref}
\include{configs} 

\newcommand{\dn}{\delta \mathcal{N}}
\newcommand{\fnl}{f_{\rm NL}}

\title{A separate universe view of the asymmetric sky}
\author{Takeshi Kobayashi,$^{1,2}$}
\author{Marina Cort\^es,$^{2,3}$}
\author{and Andrew R Liddle$^3$}
\affiliation{$^1$Canadian Institute for Theoretical Astrophysics,
University of Toronto, \\
60 St. George Street, Toronto, Ontario M5S 3H8, Canada}
\affiliation{$^2$Perimeter Institute for Theoretical Physics, 
31 Caroline Street North,\\ Waterloo, Ontario N2L 2Y5, Canada}
\affiliation{$^3$Institute for Astronomy, University of Edinburgh, Royal
Observatory, \\
Edinburgh EH9~3HJ, United Kingdom}

\emailAdd{takeshi@cita.utoronto.ca}
\emailAdd{cortes@roe.ac.uk}
\emailAdd{arl@roe.ac.uk}
\date{\today}

\abstract{We provide a unified description of the hemispherical asymmetry in the cosmic microwave background generated by the mechanism proposed by Erickcek, Kamionkowski, and Carroll, using a $\dn$ formalism that consistently accounts for the asymmetry-generating mode throughout. We derive a general form for the power spectrum which explicitly exhibits the broken translational invariance.
This can be directly compared to cosmic microwave background observables, including the observed quadrupole and $f_{\rm NL}$ values, automatically incorporating the Grishchuk--Zel'dovich effect.
Our calculation unifies and extends previous calculations in the literature, in particular giving the full dependence of observables on the phase of our location in the super-horizon mode that generates the asymmetry. We demonstrate how the apparently different results obtained by previous authors arise as different limiting cases.
We confirm the existence of non-linear contributions to the microwave background quadrupole from the super-horizon mode identified by Erickcek et al.\ and further explored by Kanno et al., and show that those contributions are always significant in parameter regimes capable of explaining the observed asymmetry. We indicate example parameter values capable of explaining the observed power asymmetry without violating other observational bounds.}

\begin{document}
\maketitle

\section{Introduction}

There is substantial variety of work in the literature proposing
mechanisms behind the hemispherical anomaly in cosmic microwave
background (CMB) temperature first identified in WMAP data
\cite{eriksen}, and later corroborated by the {\it Planck} satellite
collaboration \cite{Ade:2013nlj} and others
\cite{Paci:2013gs,Flender:2013jja, Akrami:2014eta}. We take the observed
value of the power asymmetry to be $A = 0.06$ \cite{planckslides}, where
$2A$ is the amplitude of the power asymmetry in the direction where it
is maximal. The most popular scenario is that proposed by Erickcek et
al.~\cite{Erickcek:2008sm,Erickcek:2008jp}, which we refer to as the EKC
mechanism. This proposes that the power spectrum is modulated due to the
presence of a high-amplitude long-wavelength perturbation, the
wavelength being substantially larger than the present observable
Universe. This scenario is not viable for perturbations generated by
single-field inflaton, but can be if an auxiliary field such as a
curvaton is responsible.

More generally, cosmological models that give rise to high-amplitude field
fluctuations on long wavelengths 
can imprint distinct signatures on the small-scale density
perturbations, by breaking the translational invariance of the
background in our observable patch.
This in turn suggests the possibility of probing scales beyond our
observable Universe by searching for signs of broken translational
invariance in the sky.
A particular example scenario with such large fluctuations on long wavelengths
is proposed in~\cite{LiddleCortes},
which identifies the large-scale mode with a super-curvature mode excited by vacuum tunnelling, providing for the first time a physical framework in which complete calculations can be undertaken to test against the observed properties of perturbations including their asymmetry.

However a unifying formalism for describing the hemispherical anomaly is still missing, one that holds across different treatments in the literature, and describes the full set of effects that it imparts on CMB observables. Most of the results assume specific regimes where a given set of assumptions holds.
For this reason it is not always clear how results from different models can be compared, without repeating the calculation anew under the new assumptions, with the consequence that results in the literature don't always readily carry across different works. 

In this study we start with a field perturbation which is a combination
of the usual continuum inflationary perturbation spectrum together with
a super-horizon mode; the latter is typically inserted by hand though it
may have a physical origin as a super-curvature mode also generated by
inflation. We  proceed by treating the resulting curvature perturbation
`blindly' using the $\delta \mathcal{N}$ formalism,  i.e.\ we compute the power spectrum with a minimal set of assumptions and retain all contributions of the super-horizon mode including non-linear terms and interactions between the short- and long-wavelength modes.

Other studies in the literature also considered a global perturbation
composed of the two perturbation components
\cite{Erickcek:2008sm,Erickcek:2008jp,Dai:2013kfa,Namjoo:2013fka,Lyth:2013vha,Abolhasani:2013vaa,Kanno:2013ohv,Lyth:2014mga,Namjoo:2014nra},
some of which invoked the $\delta \mathcal{N}$ formalism,
however have not carried out an exhaustive computation keeping all linear
and non-linear terms.
This has led to some confusion in the literature regarding what observational
signals are actually induced by super-horizon field fluctuations. 
We clear up the disagreements among the literature
and unify the findings of previous studies 
by consistently incorporating all fluctuations in
the $\delta \mathcal{N}$ calculations. 
In particular, we confirm
the existence of non-linear contributions to the quadrupole first
identified in Ref.~\cite{Erickcek:2008jp} and further analyzed by Kanno
et al.~\cite{Kanno:2013ohv}, then extend those studies to include
higher-order terms. The authors of the EKC mechanism did not include
non-linear corrections to the power asymmetry or non-Gaussianity
parameter $\fnl$ in Ref.~\cite{Erickcek:2008sm}, but included these for
their quadrupole expression in Ref.~\cite{Erickcek:2008jp}. In the
regime of small non-linearities where their calculations are fully
valid, the amplitude of the generated asymmetry is too small to fit the
observed value, and so cannot be responsible for the hemispherical
anomaly, whereas our results are valid in the regime where sufficient
asymmetry can be obtained.  

Moreover, our unified treatment yields the full shape of the power
spectrum, not distinguishing between super-horizon and continuum mode
contributions to multipoles. This allows us to compare our values
directly to CMB data. In particular we can compare the derived
quadrupole and octupole with their observational bounds, avoiding
separate evaluations of the Grishchuk--Zel'dovich (GZ) \cite{GZ1978} and
EKC effects on the quadrupole as in
Ref.~\cite{Kanno:2013ohv,Lyth:2013vha}. There has been discussion in the
literature as to whether these contributions are to be regarded as
correlated or independent, particularly because the GZ effect implies a
stochastic treatment of the super-horizon mode while the EKC mechanism
presupposes an asymmetry with a well-defined direction and phase. For
this reason, in our formalism we treat the large-scale mode
non-stochastically. The observed direction of the asymmetry is generated
by a single realization of the large-scale mode and associated to a
specific phase, and we can then no longer consider its phase and
direction to be random. In particular we give the full dependence of the
different observables on the phase which the observer occupies on the
super-horizon mode.\footnote{Previous works such
as~\cite{Erickcek:2008sm,Erickcek:2008jp} also presented the phase
dependence for some quantities.}

Lastly our expressions for observables in the asymmetric sky significantly simplify those in the literature by identifying the correct background field values for evaluation. We obtain observables like the power spectrum and the non-linearity parameter, $\fnl$, by keeping their usual expressions when no anomaly is present, but evaluating them at a shifted value of the field. We find that shifting the homogeneous background value of the field to the value where we are in the large-scale perturbation correctly accounts for the effects that the asymmetric mode imparts on the sky. With this simplification we obtain the correct expressions for observables, including non-linear effects, while using standard expressions valid when no asymmetry is present. 

In Sections~\ref{sec:2} to \ref{sec:par} we use the formalism described above, treating the unified field perturbation as the source of the overall observables and keeping all terms up to second order in the field derivatives. In the regime where second-order terms are dominant, considered by Kanno et al.~\cite{Kanno:2013ohv}, we arrive at similar expressions to theirs. 
However, having found physical regimes where non-linear terms dominate we then have reason to investigate even higher-order terms as well, and we proceed to do that in Section~\ref{sec:general_calc}. There we arrive at the most general expressions for the effects the asymmetry imparts on the power spectrum, bispectrum and trispectrum, with full inclusion of non-linear terms and interactions between short- and long-wavelength modes. These expressions are new to the literature.

\section{Power spectrum with broken translational invariance}
\label{sec:2}

In this article we use the $\delta \mathcal{N}$ formalism~\cite{Starobinsky:1986fxa,Sasaki:1995aw,Wands:2000dp,Lyth:2004gb} to describe the
hemispherical asymmetry. The curvature perturbation $\zeta$ is generated by fluctuations in a field $\sigma$, typically a curvaton field in the context of the EKC mechanism \cite{Erickcek:2008sm,Erickcek:2008jp}. We are going to split the fluctuations of the field $\sigma$ into two separate contributions
$\delta \sigma$ and $\Delta \sigma$; 
later on we will identify $\delta \sigma$ as the `normal' continuum spectrum of fluctuations
sourced by the inflationary background, and use $\Delta \sigma$ to represent the
super-horizon field fluctuation invoked by the EKC mechanism. 
The curvature perturbation $\zeta$ is determined by the variation in the number of $e$-foldings at different spatial locations and can be expanded in terms of the field perturbation as
\begin{equation}
 \zeta (\boldsymbol{x}) = \mathcal{N}_\sigma 
\left( \delta \sigma (\boldsymbol{x})  + \Delta \sigma (\boldsymbol{x})
\right)
+\frac{1}{2} \mathcal{N}_{\sigma \sigma} \left( \delta \sigma
					  (\boldsymbol{x})  + \Delta
					  \sigma (\boldsymbol{x})
					 \right)^2 
+ \cdots.
\label{delta-Nzeta}
\end{equation}
Here $\mathcal{N}$ is the number of $e$-folds counted forward in time from 
an initial flat hypersurface to a final uniform-density hypersurface, and 
$\mathcal{N}_\sigma, \mathcal{N}_{\sigma \sigma}, \cdots$ denote
derivatives of $\mathcal{N}$ in terms of the value of $\sigma$
on the initial hypersurface. 
The arguments of $\mathcal{N}_\sigma, \mathcal{N}_{\sigma \sigma }, \cdots $ 
are taken as the 
background field value~$\sigma_{\mathrm{bg}}$ of the entire universe 
in the expression~(\ref{delta-Nzeta}).
Hereafter, unless explicitly stated otherwise, 
the derivatives of $\mathcal{N}$ are evaluated at the
background value~$\sigma_{\mathrm{bg}}$.
We remark that on some very large scale the 
Universe is assumed to be homogeneous and isotropic, so that the separate
universe picture is valid.

Going to the Fourier space via
\begin{equation}
 \zeta (\boldsymbol{x}) = 
\frac{1}{(2 \pi)^3} \int d^3 \boldsymbol{k} \, 
e^{i \boldsymbol{k \cdot x}} \zeta_{\boldsymbol{k}},
\end{equation}
one obtains
\begin{equation}
 \zeta_{\boldsymbol{k}} = 
\mathcal{N}_\sigma \left( \delta \sigma_{\boldsymbol{k}} + \Delta
		    \sigma_{\boldsymbol{k}}   \right)
+ 
\frac{\mathcal{N}_{\sigma \sigma}}{2 (2 \pi)^3}
\int d^3 \boldsymbol{p}
\left(
\delta \sigma_{\boldsymbol{p}} +  \Delta \sigma_{\boldsymbol{p}}
\right)
\left(
\delta \sigma_{\boldsymbol{k}-\boldsymbol{p}} +  \Delta \sigma_{\boldsymbol{k}-\boldsymbol{p}}
\right)
+ \cdots.
\end{equation}

To compute the power spectrum, we assume
$\delta \sigma$ and $\Delta \sigma$ to be uncorrelated to each other,
and also that 
the small-scale field perturbations $\delta\sigma$ are well approximated
as Gaussian, so that
\begin{equation}
  \langle \delta \sigma_{ \boldsymbol{k}} \rangle = 0,
\label{onedelta}
\end{equation}
\begin{equation}
 \langle \delta \sigma_{ \boldsymbol{k_1}} \delta \sigma_{
 \boldsymbol{k_2}} \rangle 
= (2\pi)^3 P_{\delta \sigma} (k_1) \delta ( \boldsymbol{k_1 +
k_2}),
\label{twodelta}
\end{equation}
\begin{equation}
  \langle \delta \sigma_{ \boldsymbol{k_1}} \delta \sigma_{
   \boldsymbol{k_2}} \delta \sigma_{ \boldsymbol{k_3}}  \rangle = 0,
\label{sigmabi0}
\end{equation}
where $k \equiv |\boldsymbol{k}|$.
Even though we suppose the bispectrum of~$\delta \sigma$ to vanish,
non-Gaussianities in $\zeta$ will be generated via the
non-linear relation between $\zeta$ and $\delta \sigma$.
We will see that such local-type non-Gaussianities are directly related
to the hemispherical asymmetry.
One can also study cases with other types of non-Gaussianities
arising directly from the bispectrum of the field fluctuations
themselves, by modifying the assumption~(\ref{sigmabi0}). 

Furthermore, we suppose $\Delta \sigma_{\boldsymbol{k}}$ to vanish 
at the wavenumbers on which the power spectrum will be measured. In other words, 
$\Delta \sigma_{\boldsymbol{k}}$ is assumed to be nonzero only at wavenumbers much smaller than the usually considered wavemodes. 
Then, even if $\Delta \sigma_{\boldsymbol{k}}$ is stochastically
generated, we can only observe one realization in our
current Hubble patch and thus we will treat 
$\Delta \sigma_{\boldsymbol{k}}$ as a non-stochastic quantity in the
following calculations.\footnote{Generally speaking, the large-scale
fluctuation~$\Delta \sigma$ can 
induce inhomogeneities in the power spectrum of the small-scale
fluctuations~$\delta \sigma$ through the field's self-interactions,
giving rise to translational-invariance breaking in the
right-hand side of Eq.~(\ref{twodelta}).
Furthermore, the bispectrum can evolve in time and  
thus its value does not necessarily stay zero as in Eq.~(\ref{sigmabi0}).  
For instance, in cases where $f_{\rm NL}$ is strongly scale-dependent,
the time evolution of the correlation functions of $\delta \sigma $
may become significant. 
However, for simplicity, in this article we neglect such effects and assume the 
power spectrum and bispectrum to take the forms (\ref{twodelta}) and
(\ref{sigmabi0}) at all times.}
Then we obtain
\begin{equation}
\begin{split}
 \langle \zeta_{\boldsymbol{k_1}} \zeta_{\boldsymbol{k_2}} \rangle
= &
(2 \pi)^3 \mathcal{N}_\sigma^2 \, P_{\delta \sigma}(k_1) 
\, \delta ( \boldsymbol{k_1} + \boldsymbol{k_2} )
+ \mathcal{N}_{\sigma } \mathcal{N}_{\sigma\sigma  } 
\Delta \sigma_{\boldsymbol{k_1} + \boldsymbol{k_2}} 
\left(P_{\delta \sigma} (k_1) + P_{\delta \sigma} (k_2) \right)
\\
& + \frac{\mathcal{N}_{\sigma \sigma}^2}{(2 \pi)^3}
\int d^3 \boldsymbol{p} 
\Delta \sigma_{\boldsymbol{p}} \Delta \sigma_{\boldsymbol{k_1} +
 \boldsymbol{k_2} - \boldsymbol{p}} 
P_{\delta \sigma} \left( | \boldsymbol{k_1 - p} | \right)
\\
& + \frac{\mathcal{N}_{\sigma \sigma}^2}{4 (2 \pi)^6}
\int d^3 \boldsymbol{p} \, d^3 \boldsymbol{q}
\left(
\Delta \sigma_{\boldsymbol{p}} \Delta \sigma_{\boldsymbol{q}} \Delta
\sigma_{\boldsymbol{k_1} - \boldsymbol{p}} \Delta
 \sigma_{\boldsymbol{k_2} - \boldsymbol{q}}  
+
\langle
\delta \sigma_{\boldsymbol{p}} \delta \sigma_{\boldsymbol{q}} \delta
\sigma_{\boldsymbol{k_1} - \boldsymbol{ p}} \delta \sigma_{\boldsymbol{k_2} - \boldsymbol{q}} 
\rangle
\right)
\\
& + (\mathrm{terms\, with \, \mathcal{N}_{\sigma \sigma \sigma}},
\, \mathcal{N}_{\sigma \sigma \sigma \sigma}, \, \cdots).
\label{eq6}
\end{split}
\end{equation}
Here terms that involve
$\delta(\boldsymbol{k_1})$ and $\delta(\boldsymbol{k_2})$
were dropped, as we do not consider power spectra at zero wavenumber. 


For the super-horizon fluctuation $\Delta \sigma$, let us assume
that it only has modes at a single wavenumber $\pm \boldsymbol{k}_{\rm L}$, as is often assumed and for instance is the case in the open Universe scenario proposed in Ref.~\cite{LiddleCortes}. Then we can write
\begin{equation}
  \Delta \sigma_{\boldsymbol{k}} 
= \frac{(2 \pi)^3}{2} \Delta \widetilde{\sigma} 
\left\{
\delta(\boldsymbol{k} - \boldsymbol{k}_{\rm L})
+ \delta(\boldsymbol{k} + \boldsymbol{k}_{\rm L})
\right\},
\label{eq7}
\end{equation}
where $\Delta \widetilde{\sigma}$ is real and denotes the
fluctuation amplitude. 
Two delta functions are introduced so that the fluctuation in real
space is real-valued, 
\begin{equation}
  \Delta \sigma ({\boldsymbol{x}}) = 
\Delta \widetilde{\sigma} \, \cos \left(\boldsymbol{k}_{\rm L} \cdot
				   \boldsymbol{x} \right).
\label{sigmacos}
\end{equation}
Here it should be noted that $\Delta \widetilde{\sigma}$ has no spatial dependence,
but can depend on time.
Then, taking into account that the wavenumbers under consideration 
for the power spectrum are much
larger than~$k_{\rm L}$, we arrive at
\begin{equation}
\begin{split}
\frac{1}{(2 \pi)^3} \langle \zeta_{\boldsymbol{k_1}} \zeta_{\boldsymbol{k_2}} \rangle
&  =
\mathcal{N}_\sigma^2 P_{\delta \sigma}(k_1) 
\delta (\boldsymbol{k_1} + \boldsymbol{k_2})
\\
& \! \! \! \! \! \! \! \! \! \! \! \! \! \! \! \! \! \!  
+
\mathcal{N}_\sigma \mathcal{N}_{\sigma \sigma } P_{\delta \sigma}(k_1) 
\, \Delta \widetilde{\sigma } 
\left\{
\delta (\boldsymbol{k_1} + \boldsymbol{k_2} - \boldsymbol{k}_{\rm L})
+ 
\delta (\boldsymbol{k_1} + \boldsymbol{k_2} + \boldsymbol{k}_{\rm L})
\right\}
\\
& \! \! \! \! \! \! \! \! \! \! \! \! \! \! \! \! \! \!  
+ \frac{1}{4} \mathcal{N}_{\sigma \sigma }^2 P_{\delta \sigma}(k_1) 
\, \Delta \widetilde{\sigma }^2 
\left\{
2 \delta (\boldsymbol{k_1} + \boldsymbol{k_2} )
+
\delta (\boldsymbol{k_1} + \boldsymbol{k_2} - 2 \boldsymbol{k}_{\rm L})
+
\delta (\boldsymbol{k_1} + \boldsymbol{k_2} + 2 \boldsymbol{k}_{\rm L})
\right\}.
\label{eq10}
\end{split}
\end{equation}
Here we have assumed that $P_{\delta \sigma}$ is nearly scale-invariant,
i.e., 
\begin{equation}
 P_{\delta \sigma} \left(  
\left| \boldsymbol{k_1} \pm \boldsymbol{k}_{\rm L} \right|
\right) 
=
 P_{\delta \sigma} \left( k_1 \right) .
\end{equation}
We have also neglected the 
$\langle \delta \sigma ^4 \rangle$ term in Eq.~(\ref{eq6}) as $\delta \sigma$ is small and its effect will be negligible unless $\fnl$ is huge. 
Furthermore, terms with $\mathcal{N}_{\sigma \sigma \sigma}$,
$\mathcal{N}_{\sigma \sigma \sigma \sigma }$, $\cdots$ were dropped.
Such higher-order derivatives of~$\mathcal{N}$ are neglected
hereafter, until Section~\ref{sec:general_calc} where we provide an
alternative and more general calculation.

The second and third lines in Eq.~(\ref{eq10}) show that the power spectrum of $\zeta$
does not obey translational invariance, because of the presence of the super-horizon
fluctuation~$\Delta \sigma$. 
In the limit $k_{\rm L} \to 0$, the power spectrum reduces to
\begin{align}
\langle \zeta_{\boldsymbol{k_1}} \zeta_{\boldsymbol{k_2}} \rangle
& = (2 \pi)^3
 P_{\delta \sigma}(k_1) 
\delta (\boldsymbol{k_1} + \boldsymbol{k_2})
\left(
\mathcal{N}_\sigma
+ \mathcal{N}_{\sigma \sigma} \Delta \widetilde{\sigma}
\right)^2
\label{eq12}
\\
& = (2 \pi)^3
 P_{\delta \sigma}(k_1) 
\delta (\boldsymbol{k_1} + \boldsymbol{k_2})
\left(
\left. \mathcal{N}_\sigma
\right|_{\sigma_{\mathrm{bg}} + \Delta \widetilde{\sigma}}
\right)^2,
\end{align}
where in the second line we have absorbed $\Delta \widetilde{\sigma}$
into the argument of $\mathcal{N}_\sigma$.
Thus it is clear that in this limit the super-horizon fluctuation is simply a 
homogeneous shift~$\Delta \widetilde{\sigma}$ of the background field value.
The Taylor series in Eq.~(\ref{eq12}) is taken only up to quadratic
order in~$\Delta \widetilde{\sigma}$, because we have been neglecting
third and higher derivatives of~$\mathcal{N}$ in the calculations.

The asymmetric nature of the power spectrum is clearer when moving to
real-space:
\begin{equation}
\begin{split}
\langle \zeta (\boldsymbol{x}) \zeta (\boldsymbol{y}) \rangle
&= \frac{1}{(2 \pi)^6} \int d^3 \boldsymbol{k}\, 
d^3 \boldsymbol{p} \, 
e^{i (\boldsymbol{k} \cdot \boldsymbol{x} + \boldsymbol{p} \cdot
 \boldsymbol{y})}
\langle \zeta_{\boldsymbol{k}} \zeta_{\boldsymbol{p}} \rangle
\\ 
&\supset
\frac{1}{(2 \pi)^3}  \int_{k \gg k_{\rm L}} d^3 \boldsymbol{k}\, 
e^{i \boldsymbol{k} \cdot (\boldsymbol{x} - \boldsymbol{y})}
 P_{\delta \sigma}(k) 
\left\{
\mathcal{N}_\sigma
+
\mathcal{N}_{\sigma \sigma }
\Delta \widetilde{\sigma}
\cos \left( \boldsymbol{k}_{\rm L} \cdot \boldsymbol{y}\right)
\right\}^2.
\label{power-real}
\end{split}
\end{equation}
Here $\supset$ indicates that 
we have focused on contributions to the real-space power spectrum from 
observable wavemodes, and used Eq.~(\ref{eq10}) which is an expression
valid for modes larger than~$k_{\rm L}$.  
The explicit $\boldsymbol{y}$-dependence in the
$\left\{\, \, \right\}^2$ parentheses represents
the inhomogeneity of the power spectrum, including the hemispherical asymmetry.
Note that the sinusoidal field
fluctuation Eq.~(\ref{sigmacos}) sources power asymmetry 
not only with wavenumber~$k_{\rm L}$, but also with $2 k_{\rm L}$.

\section{Cosmological observables}

The best way to observationally constrain the power spectrum with broken
translational invariance would be to do a Monte Carlo comparison with data using the spectrum in the form Eq.~(\ref{eq10}), which has not been attempted yet.
Instead of doing that, we compute the multipole moments generated
by the super-horizon fluctuations, in order to estimate the constraints
on the power spectrum.

\begin{figure}[t]
\centering
\vspace*{-1.5cm}
\includegraphics[width=12cm]{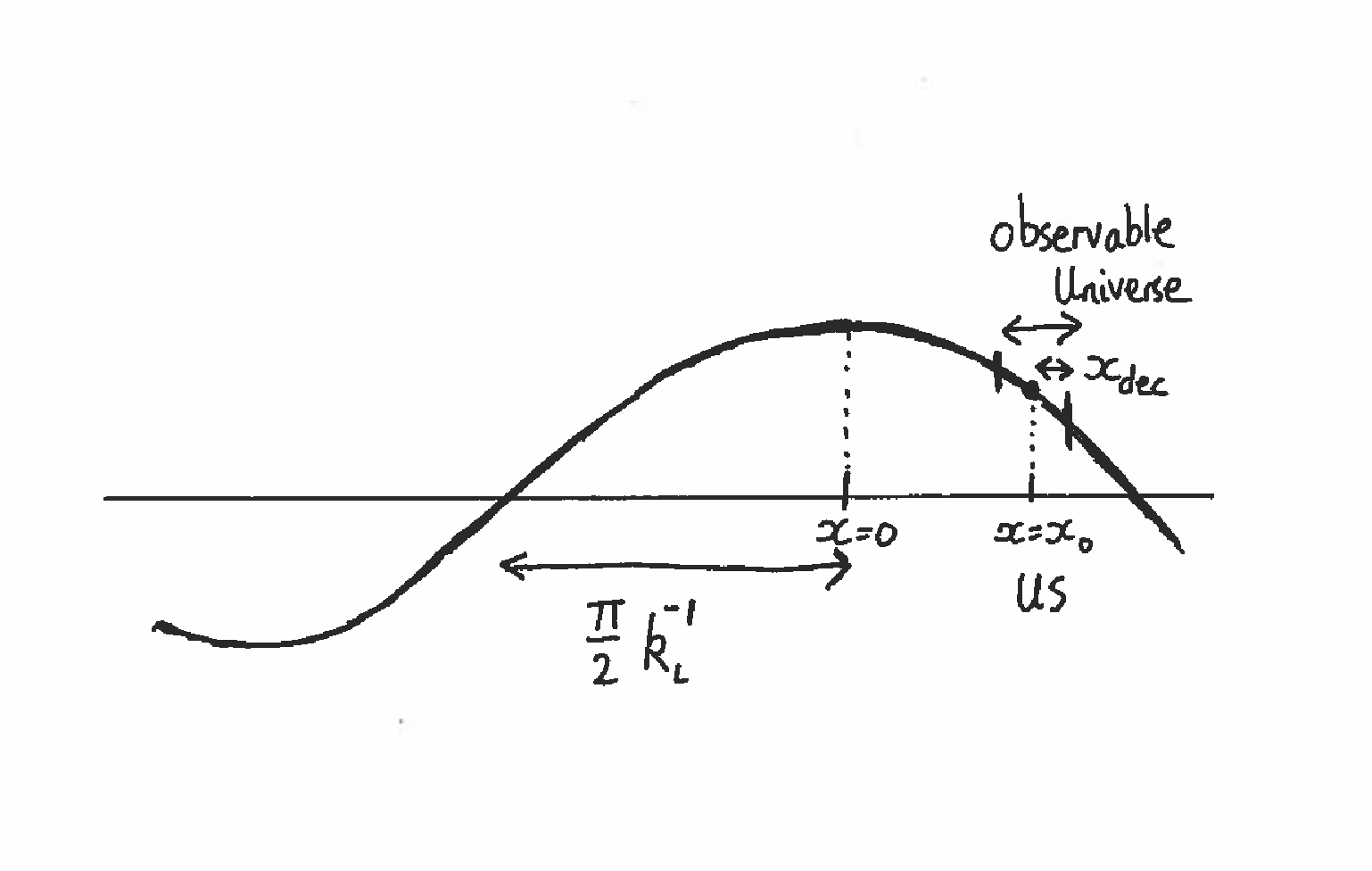}
\vspace*{-1.5cm}
\caption{The observable Universe located in the super-horizon mode. The observable patch is centered around $\boldsymbol{x}_0$ and has radius of order $x_{\rm dec}$. We assume throughout $|\boldsymbol{k}_{\rm L} \cdot (\boldsymbol{x} - \boldsymbol{x}_0)| \ll 1$ for $\boldsymbol{x}$ within our observable patch. 
}
\label{fig:setup}
\end{figure}

Let us go back to the $\delta \mathcal{N}$ expression
(\ref{delta-Nzeta}) for~$\zeta (\boldsymbol{x})$, and substitute
(\ref{sigmacos}) for $\Delta 
\sigma(\boldsymbol{x})$. 
We represent the center of our observable Universe 
by $\boldsymbol{x}_0$ and rewrite as
\begin{equation}
 \boldsymbol{k}_{\rm L} \cdot \boldsymbol{x}  = 
\boldsymbol{k}_{\rm L} \cdot (\boldsymbol{x} - \boldsymbol{x}_0)
 +  \boldsymbol{k}_{\rm L} \cdot \boldsymbol{x}_0 ,
\end{equation}
also introducing
\begin{equation}
 \theta \equiv \boldsymbol{k}_{\rm L} \cdot \boldsymbol{x}_0\,,
\end{equation}
as our particular phase in the large-scale mode, see Figure~\ref{fig:setup}. For a specific observer $\theta$ takes a fixed value, and our results will show how the observables depend on the location that the observer occupies on the super-horizon mode. As our observable patch, given by the distance to the last-scattering surface $x_{\rm dec}$, is taken to be much smaller than the wavelength of the super-horizon mode, we expand Eq.~(\ref{delta-Nzeta}) within our patch in terms of 
$|\boldsymbol{k}_{\rm L} \cdot (\boldsymbol{x} - \boldsymbol{x}_0)| \ll 1$. This
yields
\begin{equation}
\begin{split}
 \zeta (\boldsymbol{x}) = 
&
\mathcal{N}_{\sigma} 
\left\{
\delta \sigma (\boldsymbol{x})
+ \Delta \widetilde{\sigma} \cos \theta 
\right\}
 + \frac{1}{2} \mathcal{N}_{\sigma \sigma } 
\left\{
\delta \sigma (\boldsymbol{x})
+ \Delta \widetilde{\sigma} \cos \theta 
\right\}^2
\\
& -
\left\{\boldsymbol{k}_{\rm L} \cdot (\boldsymbol{x} - \boldsymbol{x}_0)\right\}
\left[
\left\{
 \mathcal{N}_{\sigma}
+ \mathcal{N}_{\sigma \sigma } 
\delta \sigma (\boldsymbol{x})
\right\}  \Delta \widetilde{\sigma} \sin \theta  
+ \frac{1}{2} \mathcal{N}_{\sigma \sigma }\Delta
\widetilde{\sigma}^2 \sin (2 \theta )
\right]
\\
& -\frac{1}{2}
\left\{\boldsymbol{k}_{\rm L} \cdot (\boldsymbol{x} -
 \boldsymbol{x}_0)\right\}^2
\left[
\left\{
\mathcal{N}_{\sigma}   
 + \mathcal{N}_{\sigma \sigma }
\delta \sigma (\boldsymbol{x})  
\right\}
\Delta \widetilde{\sigma}  \cos \theta 
+ \mathcal{N}_{\sigma \sigma }\Delta
\widetilde{\sigma}^2\cos (2 \theta )
\right]
\\
& +\frac{1}{6}
\left\{\boldsymbol{k}_{\rm L} \cdot (\boldsymbol{x} - \boldsymbol{x}_0)\right\}^3
\left[
\left\{
\mathcal{N}_{\sigma}   
+ \mathcal{N}_{\sigma \sigma }
\delta \sigma (\boldsymbol{x}) 
\right\}
  \Delta \widetilde{\sigma}  \sin \theta 
+ 2 \mathcal{N}_{\sigma \sigma }
\Delta
\widetilde{\sigma}^2 \sin (2 \theta )
\right]
\\
& + \mathcal{O} 
\left\{\boldsymbol{k}_{\rm L} \cdot (\boldsymbol{x} -
 \boldsymbol{x}_0)\right\}^4
+ (\mathrm{terms\, with \, \mathcal{N}_{\sigma \sigma \sigma}},
\, \mathcal{N}_{\sigma \sigma \sigma \sigma}, \, \cdots).
\label{zeta-mu}
\end{split}
\end{equation} 

Terms quadratic in $\boldsymbol{k}_{\rm L} \cdot (\boldsymbol{x} -
\boldsymbol{x}_0)$ source quadrupole moments in the CMB, while
cubic terms source octupole moments. The usual contributions to these
from the continuum modes are given by the $\delta \sigma$ terms in the
first line, and can be computed once the spectrum $P_{\delta \sigma}$ is
specified, while the contributions associated to the super-horizon mode
are given by the other lines in this equation.
The cross-term $\delta \sigma (\boldsymbol{x}) \Delta \widetilde{\sigma
}$ in the first line represents the shift of the average field value
within the observable patch from the background value of the
entire universe, as we will soon see.
The continuum modes contribute to each $m$-component of the multipoles, but by aligning our coordinate system to the direction of the super-horizon mode circular symmetry ensures that its contribution is only to the $m = 0$ mode of the multipoles, e.g.\ $a_{20}$ for the quadrupole.

A detailed derivation of the multipoles generated by
super-horizon perturbations that can be expressed as a Taylor
series, as in Eq.~(\ref{zeta-mu}), was performed in Ref.~\cite{Erickcek:2008jp}. 
Using their results,\footnote{In Ref.~\cite{Erickcek:2008jp}, 
multipole moments arising from a gravitational potential~$\Psi $ with a
super-horizon mode were computed. 
Their results can be applied to the 
expression~(\ref{zeta-mu}) for the curvature perturbation 
after the conversion $\Psi = -3 \zeta/5 $.} 
the super-horizon mode contribution to the quadrupole
and octupole are obtained, respectively, from the third and fourth lines of 
Eq.~(\ref{zeta-mu}) as
\begin{align}
 \Delta a_{20}  &\approx 0.10 \times
(k_{\rm L} x_{\mathrm{dec}})^2 
\left\{
\mathcal{N}_{\sigma}  \Delta \widetilde{\sigma} \cos \theta  
+ \mathcal{N}_{\sigma \sigma }
\Delta \widetilde{\sigma}^2 \cos (2 \theta )
\right\},
\label{a20}
\\
  \Delta a_{30}  &\approx
-0.019 \times
(k_{\rm L} x_{\mathrm{dec}})^3 
\left\{
\mathcal{N}_{\sigma}  \Delta \widetilde{\sigma} \sin \theta 
+ 2 \mathcal{N}_{\sigma \sigma }\Delta
\widetilde{\sigma}^2  \sin (2\theta )
\right\}.
\label{a30}
\end{align}
Here we have dropped terms proportional to 
$\mathcal{N}_{\sigma \sigma} \delta \sigma $, 
supposing they are negligible compared to $\mathcal{N}_\sigma$
terms.\footnote{In the absence of the super-horizon field
fluctuation~$\Delta \sigma$, then 
$f_{\mathrm{NL}} = (5/6) (\mathcal{N}_{\sigma \sigma } /
\mathcal{N}_{\sigma}^2)$
and 
$\zeta (\boldsymbol{x}) \simeq \mathcal{N}_{\sigma} \delta \sigma
(\boldsymbol{x})$, yielding
\begin{equation}
 \frac{\mathcal{N}_{\sigma \sigma }
\delta \sigma (\boldsymbol{x})
}{\mathcal{N}_{\sigma  }}
 \simeq \frac{6}{5} f_{\mathrm{NL}} \, \zeta (\boldsymbol{x}).
\label{Nssds}
\end{equation}
When $\left|\zeta (\boldsymbol{x}) \right|\sim 10^{-5}$, then as long as $\left|
f_{\mathrm{NL}} \right| \ll 10^5$, 
the quantity~(\ref{Nssds}) is tiny.}
Moreover, $x_{\mathrm{dec}}$ denotes the comoving distance to the
surface of last scattering. 
The super-horizon mode contribution to
the dipole cancels out 
at leading order in~$ k_{\rm L} x_{\mathrm{dec}} $
between the Sachs--Wolfe, integrated Sachs--Wolfe, and Doppler effects 
as was shown in Ref.~\cite{Erickcek:2008jp}.
Furthermore, since $\left| a_{10} \right|$ is not well
constrained due to our peculiar velocity, we do not consider the dipole
in the following discussions.
We should also remark that the numerical factors in Eqs.~(\ref{a20}) and
(\ref{a30}) have dependence on the cosmological parameters.

The terms quadratic in $\Delta \widetilde{\sigma}$ in Eqs.~(\ref{a20}) and
(\ref{a30}) arise from the non-linearities in the
conversion process of field fluctuations into curvature perturbations. 
These effects were discussed in the context of the curvaton scenario in
Ref.~\cite{Erickcek:2008jp}, and were also focused on in Ref.~\cite{Kanno:2013ohv}.

The $\left\{\, \, \right\}^2$ parentheses in the 
real-space power spectrum~(\ref{power-real}) 
can also be expanded in terms of 
$\boldsymbol{k}_{\rm L} \cdot (\boldsymbol{y}-\boldsymbol{y}_0)$, giving
\begin{multline}
\frac{1}{(2 \pi)^3}  \int_{k \gg k_{\rm L}} d^3 \boldsymbol{k}\, 
e^{i \boldsymbol{k} \cdot (\boldsymbol{x} - \boldsymbol{y})}
P_{\delta \sigma} (k)
\left( \mathcal{N}_{\sigma} + \mathcal{N}_{\sigma \sigma }
\Delta \widetilde{\sigma} \cos \theta  \right)^2
\\
\times 
\left[
1
-2 
\left\{ \boldsymbol{k}_{\rm L} \cdot (\boldsymbol{y}-\boldsymbol{y}_0) \right\}
\frac{
\mathcal{N}_{\sigma \sigma }
\Delta \widetilde{\sigma} \sin \theta 
}{
 \mathcal{N}_{\sigma} + \mathcal{N}_{\sigma \sigma }
\Delta \widetilde{\sigma} \cos \theta  
}
+ \mathcal{O} \left\{ \boldsymbol{k}_{\rm L} \cdot
	      (\boldsymbol{y}-\boldsymbol{y}_0) \right\}^2 
\right].
\label{daitenten}
\end{multline} 
Thus the all-sky averaged value of the power spectrum is obtained as
\begin{equation}
\begin{split}
 P_{\zeta} (k) 
&= P_{\delta \sigma} (k) 
\left(\mathcal{N}_\sigma + 
\mathcal{N}_{\sigma \sigma }
\Delta \widetilde{\sigma} \cos \theta 
\right)^2
\\
&=P_{\delta \sigma} (k) 
\left.
\mathcal{N}_\sigma^2
\right|_{\sigma_{\mathrm{bg}} + \Delta \widetilde{\sigma} \cos \theta }
,
\label{allsky-P}
\end{split}
\end{equation}
where the vertical line indicates evaluation at the given field value.\footnote{Here, strictly speaking the terms in the expansion~(\ref{daitenten})
with even powers of 
$\left\{ \boldsymbol{k}_{\rm L} \cdot (\boldsymbol{y}-\boldsymbol{y}_0)
\right\}$
further shift the averaged power spectrum, but we ignored such
contributions as they are suppressed by powers of $k_{\rm L}
x_{\mathrm{dec}}$.}  By expanding this last expression in terms of~$\Delta
\widetilde{\sigma}$, and dropping terms with $\mathcal{N}_{\sigma \sigma
\sigma},\, \mathcal{N}_{\sigma \sigma \sigma \sigma}, \, \cdots$, the second line reduces to the first.
The final expression in Eq.~(\ref{allsky-P}) indicates
that the average field value within our observable patch is shifted
from the background value of the entire universe 
by~$\Delta \widetilde{\sigma} \cos \theta $, at leading order in the
$k_{\mathrm{L}}$ expansion.
We stress that this shift of the mean field value needs to be taken into
account upon evaluating cosmological observables. 

The term linear in $\boldsymbol{k}_{\rm L} \cdot
(\boldsymbol{y} - \boldsymbol{y}_0)$ in~(\ref{daitenten}) gives the 
hemispherical power asymmetry. 
Half the difference of the power spectrum between spatial points that
are separated by
$ x_{\mathrm{dec}} \boldsymbol{k}_{\rm L} / k_{\rm L}$ is
\begin{equation}
\begin{split}
 A(k) \equiv \frac{1}{2} \frac{\Delta P_\zeta(k)}{P_\zeta (k)}
 &= 
-  (k_{\rm L} x_{\mathrm{dec}})
\frac{
\mathcal{N}_{\sigma \sigma }
\Delta \widetilde{\sigma} \sin \theta 
}{
\mathcal{N}_{\sigma } + 
\mathcal{N}_{\sigma \sigma }
\Delta \widetilde{\sigma} \cos \theta 
}
\\
&=
-  (k_{\rm L} x_{\mathrm{dec}})
\left.
\frac{\mathcal{N}_{\sigma \sigma } \Delta \widetilde{\sigma} \sin \theta }{\mathcal{N}_{\sigma }}
\right|_{ \sigma_{\mathrm{bg}} + \Delta \widetilde{\sigma} \cos \theta }
,
\label{Aofk}
\end{split}
\end{equation}
whose observed value is $0.06$ \cite{planckslides}.

The asymmetry induced by the super-horizon mode is generated by the non-zero value of $\mathcal{N}_{\sigma \sigma }$, which also generates cosmic non-Gaussianity. Hence the presence of the asymmetry is associated with non-Gaussianity and it is necessary to check whether the observed asymmetry is compatible with limits on non-Gaussianity. We introduce the non-linearity parameter in the observable patch,
with phase~$\theta$, by
\begin{equation}
 f_{\mathrm{NL}} (k)
= 
\left.
\frac{5}{6} \frac{\mathcal{N}_{\sigma \sigma
  }}{\mathcal{N}_{\sigma}^2} 
\right|_{ \sigma_{\mathrm{bg}} + \Delta \widetilde{\sigma } \cos
\theta }
= \frac{5}{6}
\frac{\mathcal{N}_{\sigma \sigma }}{(
\mathcal{N}_{\sigma }+ \mathcal{N}_{\sigma \sigma} \Delta
\widetilde{\sigma} \cos \theta 
)^2}.
\label{fNL}
\end{equation}
The middle expression in Eq.~(\ref{fNL}) is particularly interesting, since it lets us evaluate the non-linear parameter $\fnl$ in the  asymmetric sky with the same expression that would be used in the absence of the asymmetry, but evaluated at a shifted value of the background field. That means that we can keep the usual expression for $\fnl$, and evaluate it at the displaced value of the field, $\sigma=\sigma_{\mathrm{bg}} + \Delta \widetilde{\sigma } \cos
\theta$, according to our particular location in the long-wavelength perturbation.  The same interpretation of a usual quantity evaluated at the  displaced value of the background field also applies to the power spectrum expression in Eq.~(\ref{allsky-P}).
It should be noted that the bispectrum in the presence of the super-horizon
fluctuations should also exhibit broken translational invariance
(as for the power spectrum~(\ref{eq10})), 
and thus does not take conventional forms such as the
local-type template. 
We use $f_{\mathrm{NL}}$ defined in Eq.~(\ref{fNL}) 
simply as a rough estimate of the size of the bispectrum. 

Before ending this section, let us remark that we choose the initial
flat hypersurface for the $\delta \mathcal{N}$ calculations 
to be when a relevant wavemode
exits the Hubble horizon during inflation.
To be more specific, we focus on a certain wavenumber~$k$ for which we
consider the power asymmetry~$A(k)$, and define the time~$t_k$ as when
$k = a H$. 
The quantities in Eqs.~(\ref{allsky-P}), (\ref{Aofk}), and (\ref{fNL}), such as 
$P_{\delta \sigma}$ and $\Delta \widetilde{\sigma}$, are evaluated at $t = t_k$, 
and $\mathcal{N}_\sigma$, $\mathcal{N}_{\sigma \sigma}$, etc.\ are considered 
as derivatives in terms of $\sigma $ at $t = t_k$.
As for the multipoles, the quantities on the right-hand sides of
Eqs.~(\ref{a20}) and (\ref{a30}) can be evaluated at any time,
because the initial hypersurface can be chosen arbitrarily
for computing $\zeta$ in Eq.~(\ref{zeta-mu}).
Hence, for convenience, we also evaluate the terms in Eqs.~(\ref{a20}) and
(\ref{a30}) at $t = t_k$.

\begin{figure}[t]
\centering
\includegraphics[width=12cm]{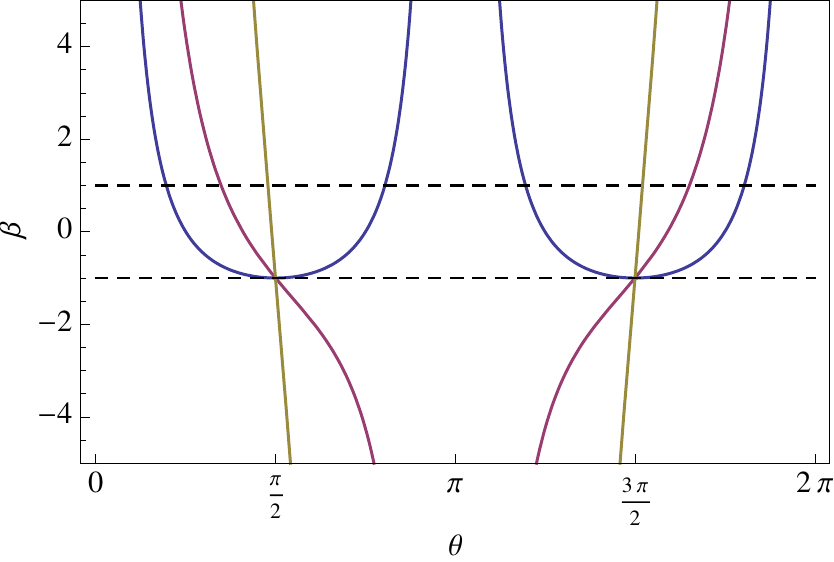}
\caption{$\beta$ as in Eq.~(4.1) evaluated for $\mathcal{N}_{\sigma}/ (\mathcal{N}_{\sigma\sigma}\Delta \widetilde{\sigma})= 0.0033$, $3$, and $30$ (blue, red, and green respectively). The dashed lines mark the regions where $|\beta|$ is or order unity.}
\label{fig:beta}
\end{figure}

\section{Generic bound on the power asymmetry and $f_{\mathrm{NL}}$}

We now derive a bound on the power asymmetry in terms of the
non-linearity parameter~$f_{\mathrm{NL}}$. 
By combining the expressions for 
the quadrupole~(\ref{a20}), power asymmetry~(\ref{Aofk}), and
non-linearity parameter~(\ref{fNL}), 
we obtain
\begin{equation}
\Delta a_{20} \, f_{\mathrm{NL}}(k) 
\approx
0.09 \times \beta A^2(k),
\qquad
\mathrm{where}
\quad
\beta  = \frac{1}{ \sin^2 \theta }
\left(
\cos 2 \theta + 
\frac{\mathcal{N}_\sigma }{\mathcal{N}_{\sigma \sigma}
\Delta \widetilde{\sigma}}
\cos \theta 
\right).
\label{eq28}
\end{equation}
This shows that to generate a significant power asymmetry, one needs large enough values of both the (super-horizon mode induced) quadrupole and the non-Gaussianity, each of which is constrained by data.
The value of $\beta$ can be much smaller than unity only when the
cosine terms are tuned to cancel each other --- see Figure~\ref{fig:beta}. 
For instance, 
when $\left|\mathcal{N}_\sigma / (\mathcal{N}_{\sigma \sigma} \Delta
\widetilde{\sigma} )\right| \ll 1$, 
choices of $\theta \approx \pi / 4, 3 \pi /4, \cdots$ suppress $\beta$ due to the 
quadrupole being suppressed, cf.~Eq.~(\ref{a20}).
Otherwise, without fine-tuning of the model parameters, $\left|
\beta \right|$ is
of order unity or larger.
Hence using $\left|\beta \right|\gtrsim 1$, we arrive at a generic bound
\begin{equation}
 \left| f_{\mathrm{NL}} (k) \right| \gtrsim 
0.09 \times
\frac{A^2(k)}{  \left| \Delta a_{20} \right|}
\gtrsim 30 \times \left(\frac{A(k)}{0.06}\right)^2 \,.
\label{fNL-A-bound}
\end{equation}
For the final expression we adopted a limit on the super-horizon mode contribution to the quadrupole of
$\left| \Delta a_{20}\right| \lesssim 1.0 \times 10^{-5}$ (c.f.\ Refs.~\cite{Erickcek:2008sm,Lyth:2013vha}); an exact constraint would require a model for the continuum contribution to the quadrupole but for our purposes requiring that this extra contribution is sub-dominant suffices.  
It is clear from Eq.~(\ref{fNL-A-bound}) that, 
in order to realize $A(k) \sim 0.06$ as suggested by {\it Planck}
\cite{planckslides}, 
$\left| f_{\mathrm{NL}} \right| $ has to be at least $\sim 30$.
The required~$f_{\mathrm{NL}}$ for producing the power asymmetry becomes
even larger when $|\beta | \gg 1$; 
then, to be compatible with the current constraints on
local-type non-Gaussianity, $f_{\rm NL} = 2.7 \pm 5.8$~\cite{Ade:2013ydc},
the non-linearity parameter needs to possess a scale-dependence.

\section{Example parameters}
\label{sec:par}

In this section, we present some example parameter sets that allow
hemispherical power asymmetry at the level suggested by observations without violating other constraints.
Such a large asymmetry is generically accompanied by 
the non-linearity parameter~$\left| f_{\mathrm{NL}} \right| $ at low
$\ell$s being larger than $\sim 30$,
cf.~Eq.~(\ref{fNL-A-bound}).
Unless this lower bound on $f_{\mathrm{NL}}$ is circumvented by
fine-tuning of the phase~$\theta$ to suppress~$\beta$,
a running of~$f_{\mathrm{NL}}$ is needed to accommodate the
CMB constraints on non-Gaussian perturbations,  but anyway the asymmetry is observed to be scale-dependent, dying away at high multipoles \cite{Ade:2013nlj,Flender:2013jja,Hirata:2009ar,Quartin:2014yaa}, and hence one should anticipate running of the non-Gaussianity in a viable model.

Let us start by summarizing the constraints that should be
satisfied. 
Firstly, we have been assuming that the 
power asymmetry is seeded by a field fluctuation with wavelength much
larger than our observable Universe, 
\begin{equation}
 k_{\rm L} x_{\mathrm{dec}} \ll 1.
\label{conA}
\end{equation}
Furthermore, upon using the 
$\delta \mathcal{N}$ formalism,
we have considered the Universe to be homogeneous and
isotropic on very large scales, thus
\begin{equation}
 \left| \mathcal{N}_\sigma \Delta \widetilde{\sigma} \right|, 
 \left| \mathcal{N}_{\sigma \sigma } \Delta \widetilde{\sigma}^2 \right|
 \ll 1.
\end{equation}
It should also be noted that we have assumed
\begin{equation}
 \left| \frac{\mathcal{N}_{\sigma \sigma }}{\mathcal{N}_{\sigma }} 
\delta \sigma (\boldsymbol{x}) \right| \ll 1
\label{conF}
\end{equation}
when computing the multipoles~(\ref{a20}) and (\ref{a30}). 
The averaged power spectrum
\begin{equation}
 P_\zeta (k ) = 
\frac{2 \pi^2}{ k^3}
\mathcal{P}_\zeta (k ),
\label{PmathcalP}
\end{equation}
whose expression is given in~(\ref{allsky-P}),
needs to satisfy the {\it Planck} normalization~\cite{Ade:2013zuv}
\begin{equation}
 \mathcal{P}_{\zeta} (k) \approx 2.2 \times  10^{-9}.
\label{conC}
\end{equation}
For the quadrupole~(\ref{a20}) and octupole~(\ref{a30}) induced by the
super-horizon field fluctuations, as described above we adopt the constraints 
\begin{equation}
 \left| \Delta a_{20} \right|, 
 \left| \Delta a_{30} \right| \lesssim 1.0 \times 10^{-5}.
\label{conE}
\end{equation}
In principle, the induced multipoles may partially cancel out the multipoles
from the continuum spectrum (i.e.\ perturbations in the absence of
super-horizon fluctuations),  and then Eq.~(\ref{conE}) would not apply, but this would be less likely than the original multipoles being coincidentally small.
We do not consider such cases in the following discussions.

The observables such as $\mathcal{P}_\zeta (k)$, $A(k)$, and
$\Delta a_{20}$ are expressed in terms of five quantities
\begin{equation}
 \mathcal{N}_{\sigma} \Delta \widetilde{\sigma}, \quad
 \mathcal{N}_{\sigma \sigma } \Delta \widetilde{\sigma}^2, \quad
k_{\rm L} x_{\mathrm{dec}}, \quad
\frac{\mathcal{P}_{\delta \sigma } (k) }{\Delta \widetilde{\sigma }^2},
\quad
\theta ,
\end{equation}
where $\mathcal{P}_{\delta \sigma}$ is related to 
the power spectrum of the field fluctuations~$P_{\delta \sigma}$ as in Eq.~(\ref{PmathcalP}).
In the following, we look into two cases, where 
$ \mathcal{N}_{\sigma} \Delta \widetilde{\sigma} $ is larger or smaller than
$\mathcal{N}_{\sigma \sigma } \Delta \widetilde{\sigma}^2$
respectively.
Let us repeat that, unless indicated otherwise, 
$\mathcal{N}_\sigma$ and $\mathcal{N}_{\sigma \sigma}$
are global universe background values, i.e.\ quantities
evaluated at~$\sigma_{\mathrm{bg}}$.

\subsection{Cases with $\left| \mathcal{N}_{\sigma} \Delta
  \widetilde{\sigma} \right| \sim \left| \mathcal{N}_{\sigma \sigma }
  \Delta \widetilde{\sigma}^2 \right| $ and $\left| \mathcal{N}_{\sigma} \Delta
  \widetilde{\sigma} \right| \gg \left| \mathcal{N}_{\sigma \sigma }
  \Delta \widetilde{\sigma}^2 \right| $}

In this case, the amplitude of the power asymmetry is given by
\begin{equation}
 A(k) \sim - (k_{\rm L} x_{\mathrm{dec}}) 
\frac{
 \mathcal{N}_{\sigma \sigma } \Delta \widetilde{\sigma}^2
}{
\mathcal{N}_{\sigma} \Delta \widetilde{\sigma}
} \sin \theta ,
\label{eq36}
\end{equation}
unless 
$\mathcal{N}_{\sigma} \Delta \widetilde{\sigma}$
and 
$\mathcal{N}_{\sigma \sigma } \Delta \widetilde{\sigma}^2$
happen to cancel each other
in the denominator of Eq.~(\ref{Aofk}).
From the assumption
$\left| \mathcal{N}_{\sigma} \Delta
  \widetilde{\sigma} \right| \gtrsim \left| \mathcal{N}_{\sigma \sigma }
  \Delta \widetilde{\sigma}^2 \right| $,
one sees that the wavelength of the super-horizon fluctuation
cannot be much larger than the size of the observable Universe if it is
to seed a power asymmetry as large as $\left|A (k)\right| \approx 0.06$.

The lower bound on $f_{\mathrm{NL}}$, Eq.~(\ref{fNL-A-bound}), is saturated
when the parameter~$\beta$ in Eq.~(\ref{eq28}) is of order unity. 
In particular, when 
$\left| \mathcal{N}_{\sigma} \Delta
  \widetilde{\sigma} \right| \gg \left| \mathcal{N}_{\sigma \sigma }
  \Delta \widetilde{\sigma}^2 \right| $, 
our observable Universe needs to be located at a special
position along the super-horizon sinusoidal fluctuation such that 
$\cos \theta$ nearly vanishes 
to have $|\beta | \lesssim 1$, cf.~Figure~\ref{fig:beta}. 

An example set of parameters that gives the observed $|A|$ while
$|f_{\mathrm{NL}}| \lesssim 10$,
enabled by tuning the phase $\theta$, 
is
\begin{equation}
\begin{split}
 &\mathcal{N}_{\sigma} \Delta \widetilde{\sigma} = 0.030, \quad
 \mathcal{N}_{\sigma \sigma } \Delta \widetilde{\sigma}^2 = 0.010 , \quad
k_{\rm L} x_{\mathrm{dec}} = 0.20, 
\\
 & \qquad \qquad
\frac{\mathcal{P}_{\delta \sigma } (k) }{\Delta \widetilde{\sigma }^2}
= 2.1 \times 10^{-6}, \quad
\theta = 1.3,
\end{split}
\end{equation}
leading to a suppressed $\beta \approx -0.06$, and yields
\begin{equation}
\begin{split}
 &\mathcal{P}_\zeta (k) \approx 2.2\times 10^{-9}, \quad
 A(k) \approx -0.06 , \quad
 f_{\mathrm{NL}}(k) \approx 8, 
\\
& \qquad 
\Delta a_{20}  \approx - 2.3 \times 10^{-6}, \quad
\Delta a_{30}  \approx - 6 \times 10^{-6}.
\end{split}
\end{equation}
Furthermore, if the typical amplitude of $\delta \sigma (\boldsymbol{x})$ is given by
the field's power spectrum as 
$\delta \sigma (\boldsymbol{x}) \sim \mathcal{P}_{\delta \sigma}^{1/2} (k)$,
then the constraint Eq.~(\ref{conF}) is also satisfied.

\subsection{Case with $\left| \mathcal{N}_{\sigma} \Delta
  \widetilde{\sigma} \right| \ll \left| \mathcal{N}_{\sigma \sigma }
  \Delta \widetilde{\sigma}^2 \right| $}
\label{subsec:caseB}

Here we study the limit considered by Kanno et al~\cite{Kanno:2013ohv}, which focused on the
non-linearity in the field fluctuations.  When the quadratic contribution~$\mathcal{N}_{\sigma \sigma} \Delta
\widetilde{\sigma}^2$ dominates over the linear~$\mathcal{N}_{\sigma} \Delta
\widetilde{\sigma}$, 
then unless $\cos \theta$ is close to zero,
the power asymmetry is directly tied to the wavenumber of
the super-horizon mode,
\begin{equation}
 A(k) \simeq -  (k_{\rm L} x_{\mathrm{dec}}) \tan \theta ,
\end{equation}
and $\beta$ is set merely by the phase, i.e., $\beta \sim \cos 2 \theta / \sin^2 \theta$.
In this limit, our relation (\ref{eq28}) simplifies as
(ignoring phase factors, hence use of $\sim$),
\begin{equation}
 \left| \Delta a_{20}\, f_{\mathrm{NL}} \right|
\sim 3 \times 10^{-4} \left( \frac{A}{0.06} \right)^2\, ,
\end{equation}
agreeing with Eq.~(7) of
Ref.~\cite{Kanno:2013ohv}.\footnote{Ref.~\cite{Kanno:2013ohv} studied
a case where two fields contribute to the curvature
perturbations. However, there it was assumed that only one of the two
fields gives non-linear contributions, and thus their Eq.~(7) matches
with our single-field result.} 
A similar expression was also obtained in the context of the curvaton
scenario in Eq.~(10) of Ref.~\cite{Erickcek:2008sm},
though it should be noted that the parameter~$A$ in Ref.~\cite{Erickcek:2008sm} is
defined as $\Delta P_{\zeta} / P_\zeta$, i.e. twice the $A$ used in this
article. 

Now, $|\beta| \lesssim 1$ is realized in a rather large range of phase
values around $\theta \sim \pi / 2$ and $3 \pi / 2$, 
as can be seen in Figure~\ref{fig:beta}.
For instance, a set of parameters
that saturates the lower bound~(\ref{fNL-A-bound}) is
\begin{equation}
\begin{split}
 &\mathcal{N}_{\sigma} \Delta \widetilde{\sigma} = 0.0010, \quad
 \mathcal{N}_{\sigma \sigma } \Delta \widetilde{\sigma}^2 =  0.15, \quad
k_{\rm L} x_{\mathrm{dec}} = 0.040,
\\
 & \qquad \qquad 
\frac{\mathcal{P}_{\delta \sigma } (k) }{\Delta \widetilde{\sigma }^2}
= 3.3 \times 10^{-7}, \quad
\theta = 1.0,
\end{split}
\end{equation}
giving $\beta \approx -0.6$, and 
\begin{equation}
\begin{split}
 &\mathcal{P}_\zeta (k) \approx 2.2 \times 10^{-9}, \quad
 A(k) \approx -0.06 , \quad
 f_{\mathrm{NL}}(k) \approx 19, 
\\
& \qquad 
\Delta a_{20}  \approx - 1.0 \times 10^{-5}, \quad
\Delta a_{30}  \approx - 3 \times 10^{-7}.
\end{split}
\end{equation}
If $\delta \sigma (\boldsymbol{x}) \sim \mathcal{P}_{\delta
\sigma}^{1/2} (k)$ 
further holds, then the constraint Eq.~(\ref{conF}) is also satisfied.

However, the condition
$\left| \mathcal{N}_{\sigma} \Delta
  \widetilde{\sigma} \right| \ll \left| \mathcal{N}_{\sigma \sigma }
  \Delta \widetilde{\sigma}^2 \right| $
may be suggesting the breakdown of the Taylor expansion of
$\mathcal{N}$ for 
$\sigma  = \sigma_{\mathrm{bg}} \pm \Delta \widetilde{\sigma}$. 
If this is actually the case, then the computation ignoring
terms with~$\mathcal{N}_{\sigma \sigma \sigma }$ and
higher-order derivatives is invalid, motivating a more general calculation that we undertake in the next section.

\section{Full account of all non-linearities}
\label{sec:general_calc}

Until now we have expanded (derivatives of) the
$e$-folding number~$\mathcal{N}$ in terms of $\Delta \sigma
(\boldsymbol{x})$ to carry out the integrations. 
However the expansion can break down
when the super-horizon field fluctuation~$\Delta \sigma $ is
much larger than the small-scale fluctuations~$\delta \sigma$,
and there does not exist a hierarchy 
of $\left| \mathcal{N}_\sigma \Delta \widetilde{\sigma} \right| \gg 
\left| \mathcal{N}_{\sigma \sigma} \Delta \widetilde{\sigma}^2 \right| \gg 
\left| \mathcal{N}_{\sigma \sigma \sigma} \Delta \widetilde{\sigma}^3
\right| \gg 
\cdots $.
Furthermore, we saw in Section~\ref{sec:par} that such a hierarchy is
typically absent in parameter regimes capable of explaining the observed asymmetry. 
In this section we generalize our treatment by avoiding $\Delta \sigma$ expansions,
instead expanding quantities directly in terms of the
wavemode of the super-horizon fluctuations~$k_{\rm L}$ (in units of the inverse of the horizon distance $x_{\rm dec}$).
The $k_{\rm L}$ expansions should be valid whenever considering fluctuations with
wavelengths much larger than the size of the observable Universe. 

The curvature perturbation is expressed 
in terms of the $\delta\mathcal{N}$ formalism as
\begin{equation}
 \zeta (\boldsymbol{x}) = 
\left. \mathcal{N}\right|_{\sigma (\boldsymbol{x})}
-
\left.
 \mathcal{N}\right|_{\sigma_{\rm bg}},
\label{delta-N-app}
\end{equation}
where the field value 
at each spatial position on the initial flat hypersurface is written as
\begin{equation}
 \sigma (\boldsymbol{x} ) = \sigma_{\mathrm{bg}} + \delta \sigma
  (\boldsymbol{x}) + \Delta \sigma (\boldsymbol{x}).
\end{equation}
While $\Delta \sigma$ may be large, we consider the perturbative
expansion in terms of the continuum 
spectrum of fluctuations~$\delta \sigma$ to be valid.
Thus we first expand $\zeta$ in terms of~$\delta \sigma$,
\begin{equation}
\begin{split}
 \zeta (\boldsymbol{x}) 
& = 
\left. \mathcal{N}\right|_{\sigma_{\mathrm{bg}} + \Delta
 \sigma (\boldsymbol{x})} 
- \left. \mathcal{N}\right|_{\sigma_{\rm bg}}
\\
& \quad 
+ \delta \sigma (\boldsymbol{x}) \, 
\left. 
\mathcal{N}^{(1)}\right|_{\sigma_{\mathrm{bg}} + \Delta \sigma
(\boldsymbol{x})} 
+ \frac{\delta \sigma^2 (\boldsymbol{x}) }{2} 
\left.
\mathcal{N}^{(2)}\right|_{\sigma_{\mathrm{bg}} + \Delta \sigma
(\boldsymbol{x})}
+ \cdots,
\label{zetaofx}
\end{split}
\end{equation}
where the $n$th derivative of~$\mathcal{N}$ in terms of the
field~$\sigma$ is denoted as~$\mathcal{N}^{(n)}$ in this section. 
Note that the $e$-folding derivatives are evaluated at a field value shifted by $\Delta \sigma$.

As in the earlier treatment, 
to compute the power spectrum
we treat the fluctuation~$\Delta \sigma $ 
as a non-stochastic quantity with the sinusoidal form,
\begin{equation}
   \Delta \sigma ({\boldsymbol{x}}) = 
\Delta \widetilde{\sigma} \, \cos \left(\boldsymbol{k}_{\rm L} \cdot
				   \boldsymbol{x} \right),
\end{equation}
and also adopt the correlation relations 
(\ref{onedelta}), (\ref{twodelta}), and (\ref{sigmabi0}) for~$\delta \sigma $.
Then the power spectrum of the Fourier modes of~$\zeta$ can be written as
\begin{equation}
\begin{split}
\langle \zeta_{\boldsymbol{k}_1}
\zeta_{\boldsymbol{k}_2} \rangle
&
= 
\int d^3\boldsymbol{x} \, d^3\boldsymbol{y} \, 
e^{-i (\boldsymbol{k}_1 \cdot \boldsymbol{x} + \boldsymbol{k}_2 \cdot
\boldsymbol{y})} 
\Biggl[
\left(
\left. \mathcal{N} \right|_{\sigma_{\mathrm{bg}} + \Delta \sigma
(\boldsymbol{x})}
- \left. \mathcal{N}\right|_{\sigma_{\rm bg}}
\right)
\left(
\left. \mathcal{N} \right|_{\sigma_{\mathrm{bg}} + \Delta \sigma
(\boldsymbol{y})}
- \left. \mathcal{N}\right|_{\sigma_{\rm bg}}
\right)
\\
& \qquad \qquad \qquad \qquad \qquad 
 + \frac{1}{2} \left(
\left. \mathcal{N} \right|_{\sigma_{\mathrm{bg}} + \Delta \sigma
(\boldsymbol{x})}
- \left. \mathcal{N}\right|_{\sigma_{\rm bg}}
\right)
\left. \mathcal{N}^{(2)} \right|_{\sigma_{\mathrm{bg}} + \Delta \sigma
(\boldsymbol{y})}
\langle \delta \sigma^2 (\boldsymbol{y})
\rangle
\\
& \qquad \qquad \qquad \qquad \qquad 
 + \frac{1}{2} \left(
\left. \mathcal{N} \right|_{\sigma_{\mathrm{bg}} + \Delta \sigma
(\boldsymbol{y})}
- \left. \mathcal{N}\right|_{\sigma_{\rm bg}}
\right)
\left. \mathcal{N}^{(2)} \right|_{\sigma_{\mathrm{bg}} + \Delta \sigma
(\boldsymbol{x})}
\langle \delta \sigma^2 (\boldsymbol{x})
\rangle
\\
& \qquad \qquad \qquad \quad \, \, \, \, \, 
 + \left. \mathcal{N}^{(1)} \right|_{\sigma_{\mathrm{bg}} + \Delta \sigma
(\boldsymbol{x})}
\left. \mathcal{N}^{(1)} \right|_{\sigma_{\mathrm{bg}} + \Delta \sigma
(\boldsymbol{y})}
\langle \delta \sigma (\boldsymbol{x})\, 
\delta \sigma (\boldsymbol{y}) \rangle
+ \mathcal{O} (\delta \sigma)^4
\Biggr].
\label{eq6.5}
\end{split}
\end{equation}

Now, unlike in the previous sections, 
we directly expand quantities in terms of~$ \boldsymbol{k}_{\rm L} \cdot
(\boldsymbol{x} - \boldsymbol{x}_0)$,
without carrying out $\Delta \sigma $~expansions.
This gives, for the derivatives of~$\mathcal{N}$, 
\begin{equation}
\begin{split}
\left.
 \mathcal{N}^{(n)} 
\right|_{\sigma_{\mathrm{bg}} + \Delta \sigma (\boldsymbol{x})}
&= 
\mathcal{N}^{(n)}_\theta 
- \boldsymbol{k}_{\rm L} \cdot (\boldsymbol{x} - \boldsymbol{x}_0) \, 
\mathcal{N}^{(n+1)}_\theta  \Delta \widetilde{\sigma} \sin \theta 
\\
&
\! \! \! \! \! \! \! \! \! \! \! \! \! \! \! \! \! \! \! \! \! \! \! \! 
\! \! \! \! \! \! \! \! \! \! \! \! \! \! \! \!
- \frac{\left\{\boldsymbol{k}_{\rm L} \cdot (\boldsymbol{x} - \boldsymbol{x}_0)
\right\}^2}{2} 
\left(
\mathcal{N}^{(n+1)}_\theta  \Delta \widetilde{\sigma} \cos \theta 
- \mathcal{N}^{(n+2)}_\theta  \Delta \widetilde{\sigma}^2 \sin^2 \theta 
\right)
\\
&
\! \! \! \! \! \! \! \! \! \! \! \! \! \! \! \! \! \! \! \! \! \! \! \! 
\! \! \! \! \! \! \! \! \! \! \! \! \! \! \! \!
+ \frac{\left\{\boldsymbol{k}_{\rm L} \cdot (\boldsymbol{x} - \boldsymbol{x}_0)
\right\}^3}{6} 
\left(
\mathcal{N}^{(n+1)}_\theta  \Delta \widetilde{\sigma} \sin \theta 
+ 3 \mathcal{N}^{(n+2)}_\theta  \Delta \widetilde{\sigma}^2 \sin \theta 
\cos \theta - 
\mathcal{N}_\theta^{(n+3)} \Delta \widetilde{\sigma}^3 \sin^3 \theta 
\right)
\\
&
\! \! \! \! \! \! \! \! \! \! \! \! \! \! \! \! \! \! \! \! \! \! \! \! 
\! \! \! \! \! \! \! \! \! \! \! \! \! \! \! \!
+ \mathcal{O}
\left\{\boldsymbol{k}_{\rm L} \cdot (\boldsymbol{x} - \boldsymbol{x}_0)
\right\}^4.
\label{eq6.6}
\end{split}
\end{equation}
Here $ \mathcal{N}^{(n)}_{\theta } $ is an abbreviation for 
\begin{equation}
 \mathcal{N}^{(n)}_{\theta } \equiv
\left.
 \mathcal{N}^{(n)}
\right|_{\sigma_{\mathrm{bg}} + \Delta \widetilde{\sigma} \cos
  \theta},
\end{equation}
which is an $\boldsymbol{x}$-independent quantity.
Expression~(\ref{eq6.6}) can be Fourier expanded using
\begin{equation}
 \frac{1}{(2 \pi )^3}
\int d^3 \boldsymbol{x}\, 
e^{-i \boldsymbol{k}\cdot \boldsymbol{x}} 
\left\{ \boldsymbol{k}_{\rm L} \cdot (\boldsymbol{x} - \boldsymbol{x}_0)
\right\}^n
= 
e^{-i \boldsymbol{k} \cdot \boldsymbol{x}_0}
\left( i \boldsymbol{k}_{\rm L} \cdot \nabla_{\boldsymbol{k}} \right)^n 
\delta (\boldsymbol{k}),
\label{nabla-delta}
\end{equation}
yielding
\begin{equation}
\begin{split}
\mathcal{N}^{(n)}_{\boldsymbol{k}} 
&= \frac{1}{(2\pi )^3}
\int d^3 \boldsymbol{x} \, 
e^{-i \boldsymbol{k \cdot x}}
\left.
\mathcal{N}^{(n)} 
\right|_{\sigma_{\mathrm{bg}} + \Delta \sigma (\boldsymbol{x})}
\\
&
=
 \mathcal{N}_\theta^{(n)} \delta(\boldsymbol{k}) 
-  \mathcal{N}_\theta^{(n+1)} 
\Delta \widetilde{\sigma} \sin \theta \, 
e^{-i \boldsymbol{k} \cdot \boldsymbol{x}_0}
i \boldsymbol{k}_{\rm L} \cdot \nabla_{\boldsymbol{k}} 
\delta(\boldsymbol{k}) 
\\
& \quad
 -\frac{1}{2} \left(
\mathcal{N}^{(n+1)}_\theta  \Delta \widetilde{\sigma} \cos \theta 
- \mathcal{N}^{(n+2)}_\theta  \Delta \widetilde{\sigma}^2 \sin^2 \theta 
\right)
e^{-i \boldsymbol{k} \cdot \boldsymbol{x}_0}
(i \boldsymbol{k}_{\rm L} \cdot \nabla_{\boldsymbol{k}} )^2
\delta(\boldsymbol{k}) 
 + \mathcal{O}(k_{\rm L})^3.
\label{Nkofn}
\end{split} 
\end{equation}
Here we have denoted terms that contain cubic or higher orders of
$\boldsymbol{k}_{\rm L}$ by $\mathcal{O}(k_{\rm L})^3$.
In terms of~$\mathcal{N}^{(n)}_{\boldsymbol{k}} $,
the power spectrum~(\ref{eq6.5}) can be written as
\begin{equation}
\begin{split}
 \langle \zeta_{\boldsymbol{k}_1}
\zeta_{\boldsymbol{k}_2} \rangle
&= 
(2\pi)^6 
\left\{ 
\mathcal{N}_{\boldsymbol{k}_1} - \left. \mathcal{N} \right|_{\sigma_\mathrm{bg}} \delta
(\boldsymbol{k}_1) 
 \right\}
\left\{ 
\mathcal{N}_{\boldsymbol{k}_2} - \left. \mathcal{N} \right|_{\sigma_\mathrm{bg}} \delta
(\boldsymbol{k}_2) 
 \right\}
\\
& 
\! \! \! \! \! \! \! \! \! \! \! \! \! \! \! \! \! \! \!
+ \frac{(2 \pi)^3}{2}
\int d^3 \boldsymbol{p}\,  P_{\delta \sigma} (p)
\left[
\mathcal{N}_{\boldsymbol{k}_1}^{(2)}
\left\{ 
\mathcal{N}_{\boldsymbol{k}_2} - \left. \mathcal{N}\right|_{\sigma_\mathrm{bg}} \delta
(\boldsymbol{k}_2) 
 \right\}
+
\mathcal{N}_{\boldsymbol{k}_2}^{(2)}
\left\{ 
\mathcal{N}_{\boldsymbol{k}_1} - \left. \mathcal{N}\right|_{\sigma_\mathrm{bg}} \delta
(\boldsymbol{k}_1) 
 \right\}
\right]
\\
& \! \! \! \! \! \! \! \! \! \! \! \! \! \! \! \! \! \! \!
+ (2\pi )^3 \int d^3 \boldsymbol{p} \, 
P_{\delta \sigma} (p)
\mathcal{N}_{\boldsymbol{k}_1 - \boldsymbol{p}}^{(1)}
\mathcal{N}_{\boldsymbol{k}_2 + \boldsymbol{p}}^{(1)}
+ \mathcal{O}(\delta \sigma)^4.
\label{eq6.10}
\end{split}
\end{equation}
Here we note that Eq.~(\ref{nabla-delta}) vanishes for
$\boldsymbol{k} \neq \boldsymbol{0}$, and so
does~$\mathcal{N}^{(n)}_{\boldsymbol{k}} $ 
up to finite order in~$\boldsymbol{k}_{\rm L}$.  
Thus, focusing on nonzero wavenumbers
\mbox{$\boldsymbol{k}_1, \boldsymbol{k}_2 \neq \boldsymbol{0}$},
the first and second lines on the right-hand side of Eq.~(\ref{eq6.10}) can be dropped.
After carrying out integration by parts on the third line, 
and ignoring the scale-dependence of the power spectrum of the field
fluctuations by assuming
\begin{equation}
0 = 
\left.
 \left( i \boldsymbol{k}_{\rm L} \cdot \nabla_{\boldsymbol{p}}\right)^m
P_{\delta \sigma} 
\left( \left| \boldsymbol{k} - \boldsymbol{p} \right| \right)
\right|_{\boldsymbol{p} = \boldsymbol{0}}
\qquad
\mathrm{for}
\quad 
m  = 1,2, \cdots,
\end{equation}
one finds
\begin{equation}
\begin{split}
  \langle \zeta_{\boldsymbol{k}_1}
\zeta_{\boldsymbol{k}_2} \rangle
= &
\frac{(2 \pi)^3}{2}
\left\{ P_{\delta \sigma} (k_1) + P_{\delta \sigma} (k_2) \right\}
\Biggl[
\left(\mathcal{N}_\theta^{(1)}\right)^2 \delta (\boldsymbol{k})
\\
& \qquad \qquad
- 2 \mathcal{N}^{(1)}_\theta \mathcal{N}^{(2)}_\theta
\Delta \widetilde{\sigma} \sin \theta \, 
e^{-i \boldsymbol{k}\cdot \boldsymbol{x}_0}
(i \boldsymbol{k}_{\rm L} \cdot \nabla_{\boldsymbol{k}})
\delta(\boldsymbol{k})
\\
& 
- \left\{
\mathcal{N}^{(1)}_\theta \mathcal{N}^{(2)}_\theta 
\Delta \widetilde{\sigma} \cos \theta -
\left( 
\mathcal{N}^{(1)}_\theta \mathcal{N}^{(3)}_\theta  
+
\left(\mathcal{N}^{(2)}_\theta\right)^2
 \right)
\Delta \widetilde{\sigma}^2 \sin^2 \theta 
\right\}
\\
&
\qquad \qquad \qquad \qquad \qquad \qquad \qquad \qquad 
\times
e^{-i \boldsymbol{k}\cdot \boldsymbol{x}_0}
(i \boldsymbol{k}_{\rm L} \cdot \nabla_{\boldsymbol{k}})^2
\delta(\boldsymbol{k})
\left.
\Biggr]
\right|_{\boldsymbol{k} = \boldsymbol{k}_1 + \boldsymbol{k}_2}
\\
&
+ \mathcal{O}(k_{\rm L})^3 + \mathcal{O}(\delta \sigma)^4 
.
\label{zetaknew}
\end{split}
\end{equation}
The terms with derivatives of the delta function represent the
broken translational invariance. 

Hereafter we neglect terms of $\mathcal{O}(\delta \sigma)^4$ and
$\mathcal{O}(k_{\rm L})^3$.
Also ignoring contributions from zero wavemodes,
the power spectrum in real space is computed by
integrating Eq.~(\ref{zetaknew}), giving
\begin{equation}
\begin{split}
\! \! \!
 \langle \zeta ({\boldsymbol{x}})
\zeta ({\boldsymbol{y}}) \rangle 
=&
\frac{1}{(2 \pi)^3}\int d^3 \boldsymbol{k} \, 
e^{i \boldsymbol{k} \cdot (\boldsymbol{x} - \boldsymbol{y})}
P_{\delta \sigma} (k)
\Biggl[
\left( \mathcal{N}^{(1)}_\theta \right)^2 - 
2 
\boldsymbol{k}_{\rm L} \cdot (\boldsymbol{y} - \boldsymbol{x}_0) \, 
\mathcal{N}^{(1)}_\theta \mathcal{N}^{(2)}_\theta 
\Delta \widetilde{\sigma} \sin \theta 
\\
& 
\! \! \! \! \! \! \! \! \! 
\! \! \! \! \! \! \! \! \! \!
- \left\{ \boldsymbol{k}_{\rm L} \cdot (\boldsymbol{y} - \boldsymbol{x}_0) \right\}^2
\left\{
\mathcal{N}^{(1)}_\theta \mathcal{N}^{(2)}_\theta 
\Delta \widetilde{\sigma} \cos \theta 
- \left(
\mathcal{N}^{(1)}_\theta \mathcal{N}^{(3)}_\theta 
+ \left(\mathcal{N}^{(2)}_\theta \right)^2
\right)
\Delta \widetilde{\sigma}^2 \sin^2 \theta 
\right\}
\Biggr] .
\label{realpowerApp}
\end{split} 
\end{equation}
One can read off the all-sky averaged power spectrum, at the leading
order in $k_{\mathrm{L}}$,
\begin{equation}
 P_{\zeta} (k) = P_{\delta \sigma} (k) 
\left( \mathcal{N}_\theta^{(1)} \right)^2,
\label{PofkApp}
\end{equation}
and the hemispherical power asymmetry
\begin{equation}
 A(k) = \frac{1}{2} \frac{\Delta P_{\zeta}(k)}{P_{\zeta}(k)} 
= -  (k_{\rm L} x_{\mathrm{dec}}) 
\frac{\mathcal{N}_\theta^{(2)}
\Delta \widetilde{\sigma} \sin \theta 
}{\mathcal{N}_\theta^{(1)}}.
\label{AofkApp}
\end{equation}
We also define the non-linearity parameter as
\begin{equation}
 f_{\mathrm{NL}} (k) = \frac{5}{6}
\frac{\mathcal{N}_\theta^{(2)}}{
\left( \mathcal{N}_\theta^{(1)} \right)^2
}.
\label{fNLApp}
\end{equation}
The multipoles that arise from the super-horizon fluctuations
can be computed by expanding $\zeta (\boldsymbol{x})$ in terms of
$\boldsymbol{k}_{\rm L} \cdot (\boldsymbol{x} - \boldsymbol{x}_0)$ 
(cf.\ Eqs.~(\ref{zetaofx}) and (\ref{eq6.6}), and see also discussions around
Eqs.~(\ref{a20}) and (\ref{a30})),
\begin{align}
\! \! 
 \Delta a_{20} &\approx
0.10 \times (k_{\rm L} x _{\mathrm{dec}})^2
\left(
\mathcal{N}^{(1)}_\theta  \Delta \widetilde{\sigma} \cos \theta 
- \mathcal{N}^{(2)}_\theta \Delta \widetilde{\sigma}^2 \sin^2 \theta 
\right),
\label{a20App}
\\
\! \! 
 \Delta a_{30} &\approx
-0.019 \times
(k_{\rm L} x _{\mathrm{dec}})^3
\left(
\mathcal{N}^{(1)}_\theta  \Delta \widetilde{\sigma} \sin \theta 
+3 \mathcal{N}^{(2)}_\theta \Delta \widetilde{\sigma}^2 \sin \theta \cos \theta 
- \mathcal{N}^{(3)}_\theta  \Delta \widetilde{\sigma}^3 \sin^3 \theta 
\right).
\label{a30App}
\end{align} 

Combining the above expressions gives
\begin{equation}
\Delta a_{20} \, f_{\mathrm{NL}} (k)
 \approx
0.09 \times A^2(k)
\left(
-1 +
\frac{\mathcal{N}_\theta^{(1)}}{\mathcal{N}_\theta^{(2)} \Delta
\widetilde{\sigma }}
\frac{\cos \theta }{\sin^2 \theta }
\right).
\label{eq57}
\end{equation}
Further introducing 
\begin{equation}
 g_{\mathrm{NL}} (k) = \frac{25}{54}
\frac{\mathcal{N}_\theta^{(3)}}{
\left( \mathcal{N}_\theta^{(1)} \right)^3
}
\end{equation}
to parameterize the size of the trispectrum,
one also obtains
\begin{equation}
\Delta a_{30}\, \frac{f^3_{\mathrm{NL}}(k)}{g_{\mathrm{NL}}(k)}
\approx
0.023 \times A^3(k)
\left(
-1 
+
\frac{3 \mathcal{N}_\theta^{(2)} \cos \theta }{\mathcal{N}_\theta^{(3)}
\Delta \widetilde{\sigma} \sin^2 \theta }
+
\frac{\mathcal{N}_\theta^{(1)}}{\mathcal{N}_\theta^{(3)} \Delta
\widetilde{\sigma }^2 \sin^2 \theta }
\right),
\end{equation}
though its implication is less clear compared to Eq.~(\ref{eq57}).

The expressions in the previous sections are reproduced by 
expanding the formulae in this section by~$\Delta
\widetilde{\sigma}$ and ignoring terms of 
$\mathcal{N}^{(3)}, \mathcal{N}^{(4)}, \cdots$.
One can check that the expressions
(\ref{realpowerApp}), (\ref{PofkApp}), (\ref{AofkApp}), (\ref{fNLApp}),
(\ref{a20App}), (\ref{a30App}), (\ref{eq57})
reduce to, respectively,
(\ref{daitenten}), 
(\ref{allsky-P}), (\ref{Aofk}), (\ref{fNL}), (\ref{a20}), (\ref{a30}),
(\ref{eq28}). 

Even though we have not carried out 
$\Delta \widetilde{\sigma}$~expansions in this section,
most of the results contain terms only up to cubic order in~$\Delta
\widetilde{\sigma}$. 
This is because $\Delta \widetilde{\sigma}$
is generically accompanied by the wavenumber~$k_{\mathrm{L}}$, and thus 
terms with higher orders 
of~$\Delta \widetilde{\sigma}$ are suppressed by 
higher powers of~$k_{\mathrm{L}}$. 

\section{Summary}

We have presented a comprehensive analysis of curvature
perturbations induced by field fluctuations with wavelength larger than
our observable Universe. 
Treating the super-horizon fluctuations as non-stochastic, we evaluated
correlation functions of perturbations on small scales.
In other words, we have analyzed conditional probability distributions
of small-scale perturbations given a fixed form for the super-horizon
fluctuation.  

Our formalism improves on previous approaches by consistently
incorporating the super-horizon and small-scale fluctuations using 
the $\delta \mathcal{N}$ formalism. 
In particular, we obtained the full expression for the power spectrum 
of the curvature perturbations that
manifests translational-invariance breaking in the observable patch, 
Eqs.~(\ref{eq10}) and (\ref{zetaknew}).
We also point out the importance of choosing the appropriate
`background' value where various quantities should be evaluated, which leads to significant simplifications in the final expressions. 
Observational signatures that accompany the power
spectrum with broken translational invariance
were computed, such as the hemispherical
power asymmetry and multipole moments in the CMB, including the full dependence on the phase~$\theta$ of our location within the super-horizon mode.

Non-linear effects are introduced through the conversion process of the
field fluctuations into curvature perturbations, and also as the
perturbations source the CMB temperature fluctuations. 
As a consequence, the observational signatures show up not only at the 
wavemode~$k_{\mathrm{L}}$ of the super-horizon field
fluctuation, but also at smaller length scales.
In the first half of the article, we carried out analyses by
expanding various quantities in terms of the 
wavenumber~$k_{\mathrm{L}}$ and amplitude~$\Delta \widetilde{\sigma }$
of the super-horizon field fluctuation, 
and dropped terms with cubic or higher orders in~$\Delta \widetilde{\sigma }$. 
We then further generalized the calculations in
Section~\ref{sec:general_calc} by carrying out expansions only in terms
of~$k_{\mathrm{L}}$, but not in terms of~$\Delta \widetilde{\sigma }$,
in order to accommodate potentially large higher-order derivatives, enabling calculations such as the trispectrum
of the curvature perturbations.

Our formalism can be applied to any model where the curvature
perturbations are sourced by a field that has super-horizon fluctuations. 
We have shown that there exist parameter ranges where the 
hemispherical power asymmetry can be as large as the observed value $A \sim 0.06$,
while the non-Gaussianity is suppressed to $| f_{\mathrm{NL}} | \lesssim 10$. 
It is important to construct explicit models that realize
such parameter ranges. 
Alternatively, super-horizon fluctuations in 
models with running~$f_{\mathrm{NL}}$ may also realize significant power
asymmetry and still be consistent with existing 
constraints on non-Gaussianity. We will report on investigations of explicit models in a future publication.

We have obtained constraints on primordial power spectra with
broken translational invariance mainly through discussing the induced
multipoles in the CMB. 
However, a direct way to constrain such power spectra would be to
do a MCMC comparison with observational data using the full expression
of the spectrum presented in Eq.~(\ref{eq10}). 
It would be interesting to see how the constraint on the power
asymmetry changes, compared to current limits imposed without having
a model for the power spectrum. 
The calculations in this article can be extended to 
compute higher-order correlation functions as well. 
The higher-order functions should also exhibit translational invariance
breaking, giving rise to new shapes for the bispectrum and trispectrum. 

Spatial dependences of statistical quantities can, albeit indirectly,
give us hints about cosmological structures beyond our observable
Universe. Our formalism is useful for probing super-horizon
perturbations from observations of asymmetric signals in the sky.

\acknowledgments

M.C.\ was supported by EU FP7 grant
PIIF-GA-2011-300606 and A.R.L.\  by the Science and Technology
Facilities Council [grant numbers ST/K006606/1 and
ST/L000644/1]. Research at Perimeter Institute is supported by the
Government of Canada through Industry Canada and by the Province of
Ontario through the Ministry of Research and Innovation. A.R.L.\
acknowledges hospitality of the Perimeter Institute during part of this
work. We thank Niayesh Afshordi, Chris Byrnes, Sean Carroll, Hassan Firouzjahi,
Amir Hajian, Sugumi Kanno, Eiichiro Komatsu, David Lyth, Paolo Natoli, Donough Regan, David Seery, and Matias Zaldarriaga for discussions.

%

\end{document}